\documentclass{elsart}

\usepackage{graphics}

\usepackage{amssymb}

\begin{document}

\begin{frontmatter}


\title{Endohedal fullerenes C$_{60}$ and C$_{82}$ with silver}
\author{V.S. Gurin}

\address{Physico-Chemical Research Institute, 
           Belarusian State University, Minsk, 220080, Belarus}
\ead{E-mail: gurin@bsu.by; gurinvs@lycos.com}

\begin{abstract}
The models of endofullerenes C$_{60}$ and C$_{82}$ with silver atom
or diatomic silver are calculated with {\it ab initio} SCF Hartree-Fock methods  
including the full geometry optimization. Ag@C$_{60}$ is a bound system with
positive binding energy while Ag$_2$@C$_{60}$ is not because strong 
geometrical strain. Silver atom is located at some distance from the 
cage center in the lower-energy model, and the structure reduces
the symmetry. The endo-structures with the C$_{82}$ cage
can exist with both mono- and diatomic silver. Electronic charge transfer 
in all structures occurs from the carbon cage to silver.
\end{abstract}

\begin{keyword}
Endohedral fullerenes \sep silver \sep clusters \sep SCF calculations 

\PACS 36.40.-c \sep 61.48.+c \sep 73.22.-f
\end{keyword}
\end{frontmatter}

\section{Introduction}
\label{sec1}
Endohedral fullerenes with metal atoms and clusters are of interest 
as species in which interaction of metal and carbon atoms can be either 
of the Van der Waals nature (if size of the fullerene cage is enough 
large) or through the metal-carbon chemical bonding due to overlap of 
carbon and metal atomic orbitals \cite{Ref1,Ref2}. 
The stable C$_{60}$ molecule with closed $\pi$-electronic 
system is not very active reagent for metal atoms, the higher fullerenes 
C$_{70}$, C$_{82}$, C$_{84}$, etc. are more reactive, and possibility
to keep metal atoms and clusters evidently is more for them.
C$_{82}$ is considered as one of 'magic' higher endofullerenes \cite{C82magic}. 
The endohedral fullerenes with rare earth elements were 
produced and theoretically evaluated in many works \cite{Ref2,Ref3,Ref4},
however, silver was not considered to date for our knowledge. 
It was reported in  \cite{Ag-C60} on interaction products of silver atoms with
C$_{60}$ molecules, but no information is available on possibility of 
endo-position of silver. Meanwhile, silver atoms and small clusters 
can be stabilized in solutions and solid matrices \cite{Agn}, and 
they are of essential interest from many point of view. Silver clusters and 
nanoparticles were produced also in carbon nanotubes \cite{AgCtube1,AgCtube2} 
those can be treated as close analogs of fullerenes possessing another topology.
It should be noticed that elements with easy capability to form 
endohedral fullerenes have a low ionization energy evidencing that the ionization
process is an important step in formation of M$_n$@C$_m$, and electron
acceptor ability of fullerene is also a key factor. Silver has the 
ionization energy 7.6~eV that is very close to the known value of C$_{60}$
\cite{IEC60}. This can be a reason of different properties of endofullerenes with
silver, if they can exist, as compared with endofullerenes with active metals.
Electronic structure of these species should
be more complicated due to strong hybridization of atom orbitals (AO) of silver 
and carbon \cite{Laslo}.

In the present work, we consider a series of models built from fullerene 
molecules C$_{60}$, C$_{82}$, silver atoms and diatomic silver. 
They were calculated with {\it ab initio} SCF Hartree-Fock methods with 
full geometry optimization, and possibility of existence and some properties of
the models are evaluated.

\section{Calculation methods}
\label{sec2}

Model structures under study are displayed in Fig.~1: mono-atomic 
endo-structures and the fullerene cages with diatomic Ag$_2$.  They are 
pictured as optimized, initial positions of one Ag atom was the centre,
and for Ag$_2$@C$_{60}$ we placed two atoms along $z$-axis ($C_5$ axis for C$_{60}$
and $C_3$ for C$_{82}$). A principal possibility to embed diatomics 
may be not excluded knowing the typical interatomic distances of Ag$_2$, 
about 2.5~\AA  \cite{Ag2}, and the size of C$_{60}$ cage, 7.1~\AA.
We estimate stability through the binding energies defined with respect 
to a decay into free atoms or Ag$_2$ and empty C$_{60}$ or C$_{82}$ those
were calculated at the same level of theory.

The task on minimum energy geometries was done for ground states: 
doublets for mono-atomic endofullerenes and singlets for the models with
Ag$_2$. For empty C$_{60}$ and C$_{82}$ singlet states are 
most probable for stable species. The calculation method used was 
the self-consistent (SCF) Hartree-Fock (HF) (unrestricted for doublet states)
within the molecular orbital - 
linear combination of atomic orbitals (MOLCAO) approach and 
density functional theory (DFT) with B3LYP functional.
The basis functions were constructed with the 19-electronic effective core potential
(ECP) for Ag and the all-electronic set of STO-3G quality for carbon atoms. 
Initial coordinates of carbon atoms for C$_{60}$ were generated within 
the I$_h$ symmetry from one unique atom position (C$_5$ axis was directed as $z$),
and for C$_{82}$ they were taken from \cite{C82coord}.
The calculations were done with a NWChem 4.1-4.5 software \cite{NW}. 
The basis sets were used within this package library with no 
modifications. 
Effective charges for atoms were calculated from Mulliken occupancies 
of the optimized structures.

A final geometry was searched by the full optimization allowing any 
distortion from the initial $I_h$ or $C_{3v}$ symmetry for the structures
with C$_{60}$ and C$_{82}$, respectively, and metal atoms were
also allowed to change positions arbitrary, thus the final structures of 
minimum total energy were obtained. 

It should be noticed that the present choice of calculation method 
for the endofullerenes with silver seems not to be very simplified, 
though SCF HF may not be considered as highly adequate for many systems.
We need analyse asymmetrical clusters having 82 carbon atoms and 2 silver atoms,
and we use ECP only for silver that is commonly accepted for this metal.
Carbon atoms are treated with rather short basis set, however, the known 
results for empty $C_{60}$ are reproduced quite sufficiently, and 
the calculation with 6-31G basis is given for comparison. A full symmetry
breaking strongly enlarges the task. An account of electronic corelations
would be fruitful, however, within the framework of this paper we analyse only
geometry of ground states, and they will be considered in future for 
calculation of more properties. We present also the test calculations wit DFT
in which electronic correlation is taken into account for the selected
models, and the results are in quakitative consistence with the SCF HF method.

\begin{figure*}
\resizebox{1.0\columnwidth}{!}{%
  \includegraphics{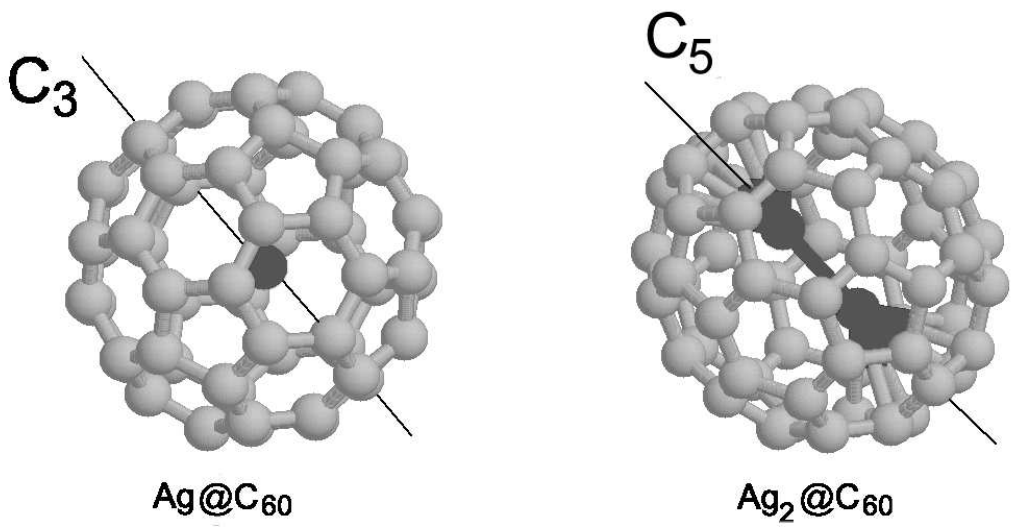}
}
\caption{Models with C$_{60}$}
\label{fig1}       
\end{figure*}

\begin{figure*}
\resizebox{1.0\columnwidth}{!}{%
  \includegraphics{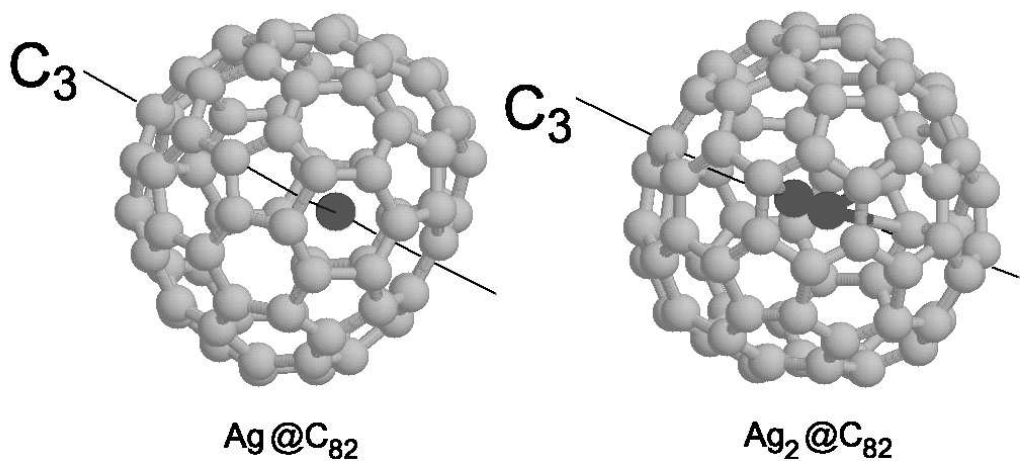}
}
\caption{Models with C$_{82}$}
\label{fig2}       
\end{figure*}

\section{Results and Discussion}
\label{sec3}

\subsection{Ag@C$_{60}$}
\label{subsec31}

The structure Ag@C$_{60}$ is quite stable according to the value of binding 
energy. The value of 10.15~eV
means that the hypothesis on Ag@C$_{60}$ has the theoretical ground.
Very recently \cite{Cu-C60} Cu@C$_{60}$ was shown to be produced 
by collision of evaporated C$_{60}$ molecules with copper plasma. As far as 
properties of copper and silver are rather close we may suggest possibility of
similar production of Ag-containing endofullerene.

The calculated Ag@C$_{60}$ appears to be noticeably deformed from the perfect initial 
geometry of C$_{60}$.
The deformation looks as the axial extension of the whole cage, 
the minimum C-C distance grows on 0.2~\AA, and the maximum one does on 0.8~\AA.
The structure losses
initial symmetry by the geometry optimization procedure, and the final one 
possesses the C$_3$ symmetry, silver atom is placed on the axis crossing the centres
of opposite  hexagons (Fig.~1). The shift of endohedral silver atom is about 
0.28~\AA. It should be remarked that a displacement of atoms and ions inside 
the C$_{60}$ cages (as well inside C$_{70}$, C$_{80}$, C$_{82}$, etc.) is 
familiar phenomenon of endofullerene science \cite{Ref12,Ref13}, and 
often this displacement appears to be even more. The shift of the endohedral
silver atom occurs approximately (but not exactly) along the C$_3$ axis, 
however, the calculation keeping the C$_3$ symmetry does not provide energy 
lower than the value calculated without symmetry (C$_1$).
Table~2 presents the numerical data for  Ag@C$_{60}$ calculated 
keeping different symmetry, and the data with additional calculation with DFT
confirm the conclusions of HF calculations. The lowest energy structures 
are not symmetrical. These geometrical features were not reported for endofullerenes 
with the C$_{60}$ cage for another metals. Likely, complicacy of silver atoms
electronic structure can be a reason of this phenomenon.

The calculated values of effective charges in the optimized clusters is $-0.1$ 
that means the interaction of silver atom is not very strong, and 
the charge transfer exists from the C$_{60}$ cage to silver atom. 
$s$-type orbitals from silver atoms contribute dominantly to highest occupied MO
(HOMO) like to the chemical bond formation in the case of alkali metals, but 
the charge transfer direction shows that silver in this case does not behave
as typical $s$-metal. The reason of this fact can be the lower energy level
of $s$-orbitals for silver than for the more active metals. In the other words, 
chemical intuitive fact of the lower electron-donor ability of silver
is consistent with these result on charge transfer.

\subsection{Ag@C$_{82}$}
\label{subsec32}

The structure {Ag@C$_{82}$ was not touched for calculations to date. 
From several isomers existing for C$_{82}$ we took the example with C$_{3v}$  
symmetry, and the arbitrary geometry distortion was allowed during calculation.
The final endo-structure appeared to be of the lower symmetry, C$_s$ only.
Silver atom distorts the cage rather strongly (Fig.~2, Table~1), but an interaction 
of Ag with carbon atoms in the optimized stable structure is weak 
since the effective charge calculated to be only -0.08 (less than 
in Ag@C$_{60}$).  
The value of binding energy amounts about 16.5~eV.  In contrast with 
Ag@C$_{60}$, in the case of silver atom within C$_{82}$ an essential 
admixing of $p$-type AOs of Ag occurs that can be a reason of the strong 
geometry distortion. 

\begin{table}
\caption{Selected interatomic distances in the optimized 
models with C$_{60}$ and C$_{82}$, \AA,  and effective charges 
at silver atoms (in units of $e$). For C$_{60}$ the results are given 
for two basis sets: STO-3G/6-31G}
\label{tab1}
\begin{tabular}{rcccccc}
\hline\noalign{\smallskip}
Model & Min C-C & Max C-C & M-M & BE, eV & Ag1 & Ag2  \\
\noalign{\smallskip}\hline\noalign{\smallskip}
C$_{60}       $ & 1.38/1.37 & 1.46/1.45 &      &       &       &        \\
Ag@C$_{60}    $ & 1.40 & 1.54 &      & 10.15   & -0.1  &        \\
Ag$_2$@C$_{60}$ & 1.36 & 1.51 & 2.37 & -24.23  & -1.34 & -1.34  \\
C$_{82}       $ & 1.34 & 1.49 &      &         &       &        \\
Ag@C$_{82}    $ & 1.36 & 1.55 &      & 16.5    & -0.08 &        \\
Ag$_2$@C$_{82}$ & 1.34 & 1.50 & 2.57 &  3.3    & -0.16 & -0.33  \\
\noalign{\smallskip}\hline
\end{tabular}
\end{table}

\begin{table}   
\caption{Comparison of total electronic energies (Hartrees) calculated 
with the two methods (SCF HF and DFT) for symmetric (Ag atom in the centre) 
and asymmetric Ag@C$_{60}$ models and coresponding 
displacement of the central atom, D, \AA}
\label{tab2}
\begin{tabular}{lcccc}
\hline\noalign{\smallskip}
Symmetry & Energy (HF) & D (HF) & Energy (DFT) & D (DFT) \\
\noalign{\smallskip}\hline\noalign{\smallskip}
$I_h$ & -2389.4572 &  0    & -2404.0437 & 0     \\
$C_1$ & -2389.4782 &  0.28 & -2404.0438 & 0.17  \\
$C_3$ & -2388.5994 &  0.01 & -2404.0435 & 0.14  \\
\noalign{\smallskip}\hline
\end{tabular}
\end{table}

\subsection{Ag$_2$@C$_{60}$}
\label{subsec33}

The calculation of binding energy for the models with diatomic silver 
cluster within C$_{60}$ (Fig.~1) does not argue on its stability. 
The interatomic distance Ag-Ag in the optimized structure is 2.37\AA
(approximately on 0.2~\AA  ~less than in bare diatomic Ag$_2$ \cite{Ag2}), 
i.e. Ag$_2$ is very strained that provides its instability within the cage,
which cannot be extended keeping C-C bonds unbroken.
C$_{60}$ cage is also subjected to distortion (Table~1).
Effective charges at metal atom are higher than in
the above structure with one silver atom. 
An analysis of energies of the frontier orbitals in the models 
with diatomic clusters within C$_{60}$ indicates that 
formation of Ag$_2$@C$_{60}$ rises both HOMO and LUMO levels. 
HOMO is contributed by $d$-type AOs of silver together with $p$-AOs 
of carbon. Thus, this model is instable mainly due to
geometrical reasons, and we may expect that larger cages can 
keep the silver clusters with more probability.

\subsection{Ag$_2$@C$_{82}$}
\label{subsec34}

In the calculations of silver diatomic clusters within C$_{82}$ 
we took also one isomer of C$_{82}$ with C$_{3v}$ symmetry and placed initially 
the Ag-Ag bond along the C$_3$ axis. The optimization method with no
symmetry restriction could relocate silver atoms to any other direction if 
structures of the lower energy appear.  Fig.~2 illustrates the final structure 
in which the Ag$_2$ cluster is shifted $\approx 0.1$~\AA  ~from the axis.
However, the deformation of C$_{82}$ cage is less than in the above case of 
Ag@C$_{82}$. It is worth to remember that one silver atom even within C$_{60}$
aspires not to be in the centre. 
The interatomic distance Ag-Ag is not much lower than in the 
bare silver diatomics (Table~1), and the value of binding energy of 
this structure is about 3.3~eV showing thus that it can exist.
HOMO of this structure is contributed mainly by $s-$AOs of silver with
admixture of $p-$AOs of silver. Thus, Ag$_2$@C$_{82}$ model can be proposed also 
as a candidate for real structures.

\section{Conclusions}
\label{sec4}

The present work with calculations of Ag$_n$@C$_{60}$ and Ag$_n$@C$_{82}$ 
($n=1,2$) models is the first attempt, for our best knowledge, of the theoretical
analysis of silver endofullerenes. 
There existed a priori feasibility to embed Ag atom and Ag$_2$ molecule
into both C$_{60}$  and C$_{82}$ from geometrical opinion. 
The calculations performed at the {\it ab initio} SCF HF level 
showed that these expectations are not too far from reality. 
The structures Ag@C$_{60}$, Ag@C$_{82}$ and Ag$_2$@C$_{82}$  
can exist, but the fullerene cages are essentially distorted from 
initial perfect symmetry, and positions of silver atoms are off-centre.
Silver does not provides electron-donor property of active metal in
all endofullerenes consideed.

\section*{Acknowledgements}

The work was performed under partial support of 
Ministry of Education of Belarus.


\begin{thebibliography}{17}

\bibitem{Ref1}
\textit {Endofullerenes. A New Family of Carbon Clusters} 
T.Akasaka and S.Nagase (Eds) (Kluwer Academic Publishers: 
      Dordrecht, 2002).
\bibitem{Ref2}
L. Forro and L. Mihaly, Rep. Prog. Phys. \textbf{64}, (2003) 649.
\bibitem{C82magic}
B.L. Zhang, C.Z. Wang, K.M. Ho, C.H. Xu, and C.T. Chan,
J. Chem. Phys. \textbf{98}, (1993) 3095.
\bibitem{Ref3}
Sh. Nagase, K. Kobayashi, and T. Akasaka, Bull. Chem. Soc. Jpn. \textbf{69},
 (1996) 2131.
\bibitem{Ref4}
J. Cioslowski and E.D. Fleischmann, J. Chem. Phys. \textbf{94},
 (1991) 3730.
\bibitem{Ag-C60}
J.A. Howard, M. Tomietto, and D.A. Wilkinson, J. Am. Chem. Soc., \textbf{113},
 (1991) 7870.
\bibitem{Agn}
A. Henglein, J. Phys. Chem. \textbf{97}, (1993) 5457.
\bibitem{AgCtube1}
J. Sloan, D.M. Wright, H.-G. Woo, S. Bailey, G. Bown, A.P.E. York, 
K.. Coleman, J.L. Huchison, and M.L.H. Geen, Chem. Commun (1999) 699.
\bibitem{AgCtube2} 
P. Corio, A.P. Santos, P.S. Santos, M.L.A. Temperini, V.W. Brar,
M.. Pimenta, and M.S. Dreselhaus, Chem. Phys. Lett. \textbf{383}, (2004) 475.
\bibitem{IEC60}
A.F. Hebard, Annu. Rev. Mater. Sci. \textbf{23}, (1993) 159.
\bibitem{Laslo}
I. Laslo, Fullerene Sci. and Technol. \textbf{1}, (1993) 11.
\bibitem{Ag2} 
M.D. Morse, Chem. Rev. \textbf{86}, (1986) 1049.
\bibitem{C82coord}
http://www.cochem2.tutkie.tut.ac.jp/Fuller/fsl/fsl.html
\bibitem{NW}
R. Harrison, J.A. Nichols, T.P. Straatsma, M. Dupuis et al. \textit{NWChem, 
A Computational Chemistry Package for Parallel Computers, Version 4.5.} 
(2004, Pacific Northwest National Laboratory, Richland, Washington 99352-0999, USA). 
\bibitem{Cu-C60}
H. Huang, M. Ata, and Y. Yoshimoto, Chem. Commun. (2004) 1206.
\bibitem{Ref12}
D.M. Poirier, M. Knupfer, and J.H. Weaver, W. Andreoni, 
K. Laasonen, M. Parrinello,  D.S. Bethune, K. Kikuchi, and Y. Achiba, 
Phys. Rev. \textbf{B49} (1994) 17403.
\bibitem{Ref13}
Y. Lian, Shangfeng Yang, and Shine Yang, J. Phys. Chem. \textbf{B106},
(2002) 3112.


\end{thebibliography}
\end{document}